\documentclass[preprint]{revtex4}

\begin{document}


\title{Giant magnetoresistance for ensembles of ferromagnetic granules in variable range hopping conductivity regime}

\author{V.I.Kozub}
\author{A.V.Shumilin}

\affiliation{A.F.Ioffe Physico-Technical Institute, St.-Petersburg
194021, Russia}

\begin{abstract}
We study an effect of moderate magnetic field on variable range
hopping conductivity in arrays of ferromagnetic granules separated
by tunnel barriers. It is shown that the resulting
magnetoresistance can be significantly larger than the standard
"giant" magnetoresistance in Fe-N-Fe-N... multilayers. The effect
is related to a gain in densities of states available for the
virtual processes  within the intermediate granules due to
magnetic-field induced alignment of the granule magnetizations.

\end{abstract}

\maketitle

\section{Introduction}

During last decades a significant interest was paid to transport
properties of metallic nanostructures which are considered as key
elements for different technologies including new nanoelectronic
devices (see, e.g., \cite{Likharev}). Important example of such
nanostructures is related to composites of metallic nanogranules
separated by dielectric (oxide) barriers. According to an extended
set of experimental data, the low-temperature conductance of such
structures exhibits the Efros-Shklovskii law $\sigma \propto \exp -
(T_0/T)^{1/2}$ typical for variable range hopping conductivity
\cite{Efros}.  The typical explanation for such a behavior was
assumed to be a correlation between sizes of the metallic granules
and of the intergranular distances \cite{Sheng}. 

Surprisingly,
only several years ago a mechanism directly relating the
temperature behavior in question to the variable range hopping (VRH) was
suggested \cite{1},\cite{2} \cite{3} . An important feature of
this mechanism was an assumption that the transport was controlled
by hops between remote granules.  The choice of corresponding pair
of granules corresponds to optimization of the sum of
intergranular tunneling action and activation exponent related to
a random potential of the structure (which is typical for VRH).
However, in contrast to the standard VRH(\cite{Efros}), the
"hopping trajectory" in this case inevitably includes some
"intermediate" granules which play a role of  "under-barrier
scatterers" like in theory of interference-mediated VRH
\cite{Spivak}.  Another important ingredient of this mechanism was
so-called co-tunneling, which included both an "elastic" channel
when the transport was supported by the single intermediate states
within the intermediate granules, and "inelastic" channel when the
transport was supported by two different electron states within at
least some of the intermediate granules (\cite{3}). 

An important
progress achieved in last years was related to a possibility to
fabricate aggregates of nearly monodisperse nanogranular
structures \cite{sample} . This progress allowed to exclude the
mechanism of (\cite{Sheng}), concentrating on the mechanism of
hopping including "intermediate" granules. In addition, the
technique exploited in (\cite{sample}) allowed to obtain
aggregates of ferromagnetic granules (in particular, on the base
of Ni, \cite{2}). One notices, that in the case of ferromagnetic
granules one expects an important effect of the spin correlations
for the hops including granules with different orientations of
magnetization. In particular, one expects that a presence of
external magnetic field leading to an alignment of magnetizations
in different granules can significantly affect the hopping
transport which results in pronounced magnetoresistance. Note that the
effect of negative magnetoresistance in ferromagnetic
nanostructures was extensively discussed starting from well-known,
effect of so-called giant magnetoresistance (GMR) (see, e.g.,
\cite{GMR}) appeared to be extremely useful for applications. The
simplest example of its experimental realization is related to
multilayers Fe-N-Fe... where Fe stands for a ferromagnetic layer
while N stands for a normal metal layer. When the magnetizations of
all Fe layers are aligned (which, in particular, can be controlled
by a relatively weak external magnetic field) an electron with a
given spin direction stays to belong to the same group of minority
or majority electrons in all Fe layers provided the spin
relaxation length is large enough. While the partial contributions
of the minority and majority groups in Fe layers are different, it
can be proved that the alignment mentioned above leads to a
decrease of total sample resistance with respect, to say, to
random orientation of different Fe layers. Among latest studies in
this direction we can mention, in particular, the papers \cite{4}
\cite{5}, \cite{6}. However, the attention was mainly paid to
tunneling transport involving  two neighboring ferromagnetic
layers or granules. In what follows we will concentrate on
variable range hopping through ensembles of ferromagnetic
granules involving co-tunneling. An important factor leading to  an
enhancement of the magnetoresistance in this case is related to
the fact that each hop involves many intermediate granules.  Thus,
the alignment of magnetizations within the intermediate granules
leads to multiplication of the effect expected for tunneling
between two neighboring granules.

The system in question is an ensemble of ferromagnetic granules
with tunnel barriers between the granules, which implies that the
charge transport through the system is due to hopping. The thin
films formed by granules of different metals with typical size 2-5
nm were obtained in our institute several years ago with a help of
laser electrodispersion (see e.g. \cite{sample}). The thickness of
the film started from monogranular layers while the character and
magnitude of the conductivity were dependent both on film
thickness and technological factors. At least, the Ni-based
structures which clearly  demonstrated Efros- Shklovskii variable
range hopping were also fabricated.

First, we will give a short discussion of hopping between granules
from normal metal, based mostly on the results of \cite{3}, then
will consider the consequences for ferromagnetic granules. Note
that the problem of variable range hopping in ensembles of
ferromagnetic granules imply many-particle effects related to
different granules. Thus to attack this problem it was necessary
to restrict ourselves by the minimal model concerning the
intergranular tunneling. In particular, we will not go into
details related to band structure of the ferromagnetic materials
except of a presence of minority and majority carriers.

\section{Variable-range co-tunneling in ensembles of normal metal
grains}

An important factor responsible for the variable-range hopping in
granular structures is related to the fact that such a hopping
inevitably involves for each hop between the remote granules a
supporting role of virtual electron states in intermediate
granules (see, e.g., \cite{1, 2}). Indeed, in contrast to standard
hopping conductivity via impurity states in semiconductors, here
there is no "direct" tunneling path - the shortest line connecting
any two granules except the neighboring ones inevitably crosses
some intermediate granules.
 In
other words, hopping in this case implies so-called co-tunneling
(elastic or inelastic - see, e.g., \cite{3}).

 For
demonstration, we will consider 3-granule array implying a hop
between granules 1 and 3 through the granule 2. Elastic
co-tunneling implies a presence of a single intermediate virtual
state. For elastic co-tunneling an electron, starting from the
initial granule 1 (virtually) occupies one of the states within
the intermediate granule (2) and then occupies the final state
within the granule 3. For the case of inelastic co-tunneling an
electron starting from granule 1 virtually occupies {\it one} of
the available states within granule 2 while {\it another} electron
starting from {\it another} state within the granule 2 finally
occupies the final state within the granule 3.

The important detail of such a co-tunneling (and of VRH
conductivity in granular materials) is related to the following.
As it can be easily estimated, for a granule with a radius of the
order of several nm the charging energy can reach $10^{-13} erg
\sim 10^3 K$. So, at low temperature most of the granules are
Coulomb blockaded. However the effect of the Coulomb blockade can
be partially lifted due to disorder potential related,  say, to
traps within the insulating barriers. Due to this disorder there
is a small number of granules with relatively small energy of
Coulomb blockade estimated with respect to "global" chemical
potential over the sample. The hopping is possible only between
these rare granules. However to reach the distant granule with
small Coulomb blockade energy the electron should pass through
virtual states (co-tunnel) on intermediate granules.

The most general expression for the tunneling amplitude between
two distant granules including both elastic and inelastic
co-tunneling can be found in \cite{3}. In our work we adopt
several simplifications that do not interfere with discussed
physics and allow us to make the expressions significantly simpler.
First, we consider only inelastic co-tunneling.  Second we
consider Coulomb blockade energies on intermediate granules to be
much larger than Coulomb blockade energies of starting and final
granules of the hop. Third, we neglect the changes in the Coulomb
gap energies of intermediate granules due to position of tunneling
electron. Finally, we neglect the difference between different
"time orderings" discussed in \cite{3}. These assumptions allow us
to ascribe some fixed Coulomb blockade energy to each intermediate
granule.  We will show that this approach allows us to catch the qualitative results
of \cite{3} and discuss the novel phenomena appearing due to the ferromagnetic nature of
granules.

Inelastic hopping process includes two states within each of the
intermediate granules $k$: the electron state $e_k$ is the state
to which the electron tunnels from the previous granule $k-1$, and
the hole state $h_k$ from which the electron tunnels to the next
granule. Let us write the contribution to the tunneling
probability from a given ensemble of states $\{e_k,h_k\}$
\begin{equation}\label{tunP0ekhk}
P\{e_k,h_k\} \propto \frac{\prod_{k=0}^{N}
f(h_k)t^2_{h_k,e_{k+1}}(1-f(e_{k+1})) }{\prod_{k=1}^{N}
{E}_{C,k}^2}  \delta\{e_k, h_k\}.
\end{equation}
Here we consider $N$ intermediate granules $1 ... N$. The index
$k=0$ stands for the initial granule of the hop and the index
$k=N+1$ is for the final granule of the hop. $t_{h_k,e_{k+1}}$ is
the tunneling matrix element between states $h_k$ and $e_{k+1}$.
$f$ is the fermi function and $E_{C,k}$ is the Coulomb gap energy
of $k$-th granule. $\delta\{e_k, h_k\}$  --- is the delta function
from the total energy determined by all the ensemble of states. It
ensures energy conservation during the hop.

To get the full tunneling probability $P$ one should sum
$P\{e_k,h_k\}$ over all possible ensembles of states
\begin{equation}\label{tunP0sum}
P = \sum_{\{e_k,h_k\}} P\{e_k,h_k\}.
\end{equation}
Without the effects of ferromagnetism, this summation yields
\begin{equation}\label{tunP1}
P \sim \frac{t^{2N} g^{2N}\epsilon_{inel}^{2N}}{E_{C}^{2N}} \exp
\frac{-\epsilon_{0,N+1}}{T}.
\end{equation}
Here $t$ is the characteristic value of tunneling matrix element
between adjusted granules, $E_C$ is the characteristic value of
Coulomb gap energy, $g$ is the density of states near the Fermi level
in a single granule, $T$ is the temperature, $\epsilon_{0,N+1}$ is
the absolute value of the difference of electron energies on
initial and final granules and $\epsilon_{inel} \sim
\epsilon_{0,N+1}/N$ is the characteristic inelastic energy. Note
that expression (\ref{tunP1}) agrees with the results of
\cite{3} for inelastic co-tunneling.

\section{Effect of ferromagnetism}

Let us now include in our consideration the ferromagnetic ordering
inside granules leading to a presence of magnetization within each
granule. We will assume that the overlapping integrals between
different neighboring granules are small enough (which is natural
for systems with hopping conductivity) and thus the intergranule
exchange interaction can be neglected. At the same time different
granules are expected to have different orientations of the
anisotropy axes and thus magnetization orientations ${\bf
n}_{{J},k}$ for different granules are different. Then, in each
granule we have different densities of states for majority ($g_M$)
and minority ($g_m$) electrons. It can be shown that the square of
the tunneling amplitude between two majority states or two
minority states in the adjusted granules is proportional to
$(1+\cos\theta_{k,k+1})/2$, where $\cos\theta_{k,k+1} = {\bf
n}_{{J},k} \cdot {\bf n}_{{J},k+1}$. The
majority$\rightarrow$minority and minority$\rightarrow$majority
tunneling amplitude squares are proportional to
$(1-\cos\theta_{k,k+1})/2$. Thus the contribution of tunneling
probability $P\{e_k,h_k\}$ acquires additional factors
corresponding to these cosines
\begin{equation}\label{tunP0ekhkH}
P\{e_k,h_k\} = P\{e_k,h_k\}^{(0)} \cdot \prod_{k=0}^{N} {\cal
C}_{k,k+1},
\end{equation}
where $P\{e_k,h_k\}^{(0)}$ is determined by expression
(\ref{tunP0ekhk}) with spin independent tunneling matrix elements
$t_{k,t+1}$, ${\cal C}_{k,k+1}$ depends on the nature (minority or
majority) of states $k$ and $k+1$:
\begin{equation}\label{calC}
\cdot {\cal C}_{k,k+1} = \left\{
\begin{array}{ll}
(1+\cos\theta_{k,k+1})/2 & M\rightarrow M, \, m\rightarrow m, \\
(1-\cos\theta_{k,k+1})/2 & m\rightarrow M, \, M\rightarrow m.
\end{array}
\right.
\end{equation}
Here $M$ stands for majority states and $m$ for minority.

The difference between the majority and minority density of states
should be taken into account in the summation procedure
(\ref{tunP0sum}). For the ferromagnetic granules this procedure
leads to the following result
\begin{equation}\label{tunP1H}
P \sim P^{(0)} \prod_{k=0}^{N} (1+ {\cal P}^2 \cos\theta_{k,k+1}),
\end{equation}
where $P^{(0)}$ is given by the expression (\ref{tunP1}) and
${\cal P} = (g_M - g_m)/(g_M+g_m)$ is the polarization inside
granules.

We assume that the conductivity of the sample is proportional to
averaged tunneling probability. So to calculate the
magnetoresistance one should average (\ref{tunP1H}) over angles
$\theta_{k,k+1}$. Note that, strictly speaking, angles
$\theta_{k,k+1}$ are not independent (although the directions
${\bf n}_{{J},k}$ are independent). However we will proceed
considering  values of $\theta_{k,k+1}$ as independent. It can be
shown that such an approach gives correct result for low fields
(when Zeeman energy $JH$ is much less than temperature) as well as
for the saturation field (when $\theta_{k,k+1} \sim 1$). With this
simplification we have
$$
G(H)/G(0) = (1+ {\cal P}^2 \left<\cos\theta
\right>)^{\overline{N}+1},
$$
where $G$ is the system conductance,  $\left<\cos\theta \right>$
is the average value of the angle between magnetic moment of
adjusted granules, $\overline{N}$ is the characteristic number of
intermediate granules. Note that for $\overline{N} = 0$ this
result reproduces the result of \cite{GranulsNoCot} where
intermediate granules were not included.

Now we will discuss the physics leading to the results obtained
above. As for material parameters, we will base on the experiments
\cite{granules3} where nanocomposites of Ni granules with a
granule size $\sim 2 nm$ (containing about 600 atoms) were
studied. The samples clearly demonstrated superparamagnetic
behavior, however, for aggregates containing about $10^3$
granules. The anisotropy constant $K$ (entering to the estimate of
anisotropy energy $KV$, where $V$ is a granule volume) according
to the experimental results can be estimated for a single granule
as $\sim 10^{6} erg/cm^3$. It was attributed to the shape
anisotropy.

The physical picture naturally depends on the
interplay between the Zeeman energy $JH$, the anisotropy energy
$KV$ and temperature $T$.
When the anisotropy energy of the granules can be neglected with
respect to the temperature or Zeeman energy (that is in relatively strong
magnetic fields corresponding for the aggregates discussed above
to fields larger than $\sim 0.1 T$) , it is possible to obtain the
following expression for averaged cosines
\begin{equation}\label{cos1}
\left<\cos\theta\right> = m^2 = \left(\frac{1}{\tanh(JH/T)} -
\frac{1}{JH/T}\right)^2.
\end{equation}
Here $m$ is the relative magnetization of the system and $J$ is
the magnetic moment of single granule.

At small fields expression (\ref{cos1}) leads to quadratic
magnetoresistance
\begin{equation}\label{MR1}
\frac{\Delta R}{R} = - {\cal P}^2 \frac{\overline{N}+1}{3}\left(
\frac{JH}{T} \right)^2,
\end{equation}
while at high fields $JH \gg T$ magnetoresistance saturates at the
value
\begin{equation}\label{MR2}
\frac{R(H)}{R} = (1+{\cal P}^2)^{-\overline{N}-1}.
\end{equation}
Note that expressions (\ref{MR1}) and (\ref{MR2}) are correct
despite the approximation of independent $\theta_{k,k+1}$.

Let us also discuss another case when anisotropy energy is strong
enough (and the Zeeman energy is relatively small), thus the
magnetic moments are always oriented along the easy anisotropy
axes. In this case it is important to know if the magnetic moments
of the granules can flip between the directions along the easy
axis. To change its direction the magnetic moment should overcome
the barrier related to the anisotropy energy ($\sim KV$). This
process occurs at the time scale $1/\tau \sim f_0 \exp(-KV/T)$
where $f_0$ is the attempt frequency. For realistic time scales of
the experiment it is possible to introduce a blocking temperature
$T_b \sim KV/25$ (see e.g. \cite{granules3}).  One notes that the
granules can flip its magnetic momentum direction only when the
temperature is larger than $T_b$ (for details see \cite{review})

Let us consider the situation of relatively strong anisotropy $KV
> T,\, JH$, however still assuming that the temperature is larger than the blocking
temperature $T_b$. (Note that basing on the results of
\cite{granules3} the blocking temperature for a single granule can
be estimated as $ < 1 K$). So the magnetic moments can flip
between the two possible directions along these axes minimizing
the Zeeman energy. In general, we have a standard situation of the
superparamagnetic particles. The relation $\left< \cos\theta
\right> = m^2$ holds in this case too, however the dependence of
$m$ on magnetic field is different (and more complex) than in the
situation considered above. At small magnetic field we obtain for
strong anisotropy
\begin{equation}\label{MR1S}
\frac{\Delta R}{R} = - {\cal P}^2 \frac{\overline{N}+1}{9}\left(
\frac{JH}{T} \right)^2,
\end{equation}
and in saturation field within the approximation of independent
angles
\begin{equation}\label{MR2S}
\frac{R(H)}{R} = (1+{\cal P}^2/4)^{-\overline{N}-1}.
\end{equation}
Thus in the case of strong anisotropy magnetoresistance has the
same form and the same dependence  on $\overline{N}$ as in the
case of weak anisotropy, but is somewhat weaker than in the case
of weak anisotropy.

The relative strength of anisotropy in the system should be
estimated with respect to the Zeeman energy as well as to the
temperature. If the temperature is larger than anisotropy energy
$T>KV$, we definitely have the case of weak anisotropy and the
magnetoresistance should be described by equations (\ref{MR1}) and
(\ref{MR2}). However at lower temperature $KV>T>T_b$  the strength
of anisotropy should be compared to the Zeeman energy. At weak
fields when Zeeman energy $JH$ is still weaker than $KV$ the
magnetoresistance is controlled by equation (\ref{MR1S}) (for
$JH<T$) and may even reach plateau given by eq. (\ref{MR2S}) for
$JH
> T$. However at stronger magnetic fields $JH \ge KV$ the
effect of the anisotropy is suppressed by magnetic field. At this
point the magnetoresistance goes up again and reaches its real
saturation (described by eq. (\ref{MR2})) for $JH \gg KV$.
Finally, when temperature is smaller than blocking temperature
$T<T_b$ the magnetization is blocked if Zeeman energy is smaller
than the anisotropy energy. At this case there is no
magnetoresistance up to the fields $JH \sim KV$. At larger fields
$JH \gg KV$ resistance saturates at the value (\ref{MR2}).

According to our estimates, the effect crucially (exponentially)
depends on the number of the intermediate granules $N$. Let
us estimate the upper boundary for this number. The hopping
character of the conductance implies that the intergranular
conductance is much smaller than $e^2/h$. Having in mind the small
size of the granules one expects that only small number of quantum
channels in granules contribute to the conductance (this number is
controlled by the cross-section of the tunneling area), we will
assume that this number is at least less than 10. Thus the hopping
character of the conductance implies that the tunneling
transparency is much less than 0.1 which, in its turn, implies
that the tunneling exponent is larger than 2. It gives the
contribution of each intergranular contact to the total tunneling
action corresponding to the hop. One has in mind that for the
Coulomb gap hopping the total tunneling action gives to the
hopping exponent  $\xi$ a contribution equal to $\xi/2$. Then, the
largest measurable value of $\xi$ is expected not to exceed 20.
Thus the total number of the intermediate granules in this case is
$10/2 - 1  = 4$. Since the exponents in Eqs.9,11 are equal to
$N+1$, in such a case one indeed expects a strong effect. Indeed,
for polarization degree 1/2  the corresponding factor is equal to
$(5/4)^5 \sim 3$. Certainly, this is rather an upper estimate of
the predicted effect.

\section{Conclusions}
To conclude, we have shown that the arrays of ferromagnetic
granules separated by tunneling barriers can in variable range
hopping regime exhibit giant magnetoresistance at least by order
of magnitude exceeding the standard magnetoresistance in
Fe-N-Fe... arrays. The origin of this magnetoresistance is related
to the fact that for random directions of the granule magnetic
moments the densities of states for the virtual processes within
the intermediate granules participating in the hop are partly
suppressed due to difference between systematics of minority and
majority electrons within different granules. Then, the external
magnetic field aligns the granule magnetic moments thus
establishing a unique systematics of majority and minority
electrons throughout the sample, thus increasing the corresponding
densities of states. The large increase of the effect for variable
range hopping conductivity is explained by a large number of the
intermediate granules participating within a single hop event, the
resulting magnetoresistance appears to be exponential in terms of
this number.

\section{Acknowledgements}

We are grateful for financial support from RFBR Foundation (Grant No. 13-02-00169).

\bibliographystyle{elsarticle-num}

\end{document}